\documentclass[conference]{IEEEtran}
\IEEEoverridecommandlockouts
\usepackage{cite} 
\usepackage{amsmath,amssymb,amsfonts,bbm} 
\usepackage{algorithmic}
\usepackage{tabularx} 
\usepackage{amsmath, bm} 
\usepackage{amsmath, autobreak} 
\usepackage[ruled, linesnumbered]{algorithm2e} 
\usepackage{subfigure} 
\usepackage{soul} 
\usepackage{balance} 
\usepackage{color} 
\usepackage{float} 
\usepackage{autobreak} 
\usepackage{graphicx} 
\usepackage{textcomp} 
\usepackage{xcolor} 
\usepackage{booktabs} 
\usepackage{multirow} 
\usepackage{bbm}
\usepackage{dsfont} 
\usepackage{footnote} 
\usepackage[]{mathtools} 
\usepackage{lipsum} 
\makesavenoteenv{algorithm} 
\usepackage{diagbox}
\usepackage{lineno}
\usepackage{tikz}
\usepackage{makecell}
\usepackage{amssymb}
\usepackage{mathrsfs}
\usepackage{calligra}
\usepackage[numbers,sort&compress]{natbib}
\usepackage[left=1.62cm,right=1.62cm,top=1.9cm]{geometry}

\usepackage[colorlinks, linkcolor=blue, anchorcolor=blue, citecolor=blue]{hyperref}

\begin{document}

\title{A Constrained Deep Reinforcement Learning Optimization for Reliable Network Slicing in a Blockchain-Secured Low-Latency Wireless Network}

\author{\IEEEauthorblockN{
Xin~Hao\IEEEauthorrefmark{1}, 
Phee~Lep Yeoh\IEEEauthorrefmark{2},~\IEEEmembership{Senior~Member,~IEEE},
Changyang~She\IEEEauthorrefmark{3},~\IEEEmembership{Senior~Member,~IEEE},
Yao Yu\IEEEauthorrefmark{4},\\
Branka~Vucetic\IEEEauthorrefmark{3},~\IEEEmembership{Life~Fellow,~IEEE},
and~Yonghui~Li\IEEEauthorrefmark{3},~\IEEEmembership{Fellow,~IEEE}
}

\IEEEauthorblockA{\IEEEauthorrefmark{1}School of Information Technology, Deakin University, Australia}
\IEEEauthorblockA{\IEEEauthorrefmark{2}School of Science, Technology and Engineering, University of the SunshineCoast, Australia}
\IEEEauthorblockA{\IEEEauthorrefmark{3}School of Electrical and Computer Engineering, The University of Sydney, Australia}
\IEEEauthorblockA{\IEEEauthorrefmark{4}School of Computer Science and Engineering, Northeastern University, China}
}

\maketitle

\begin{abstract}
Network slicing (NS) is a promising technology that supports diverse requirements for next-generation low-latency wireless communication networks. However, the tampering attack is a rising issue of jeopardizing NS service-provisioning. To resist tampering attacks in NS networks, we propose a novel optimization framework for reliable NS resource allocation in a blockchain-secured low-latency wireless network, where
trusted base stations (BSs) with high reputations are selected for blockchain management and NS service-provisioning. 
For such a blockchain-secured network, we consider that the latency is measured by the summation of blockchain management and NS service-provisioning, whilst the NS reliability is evaluated by the BS denial-of-service (DoS) probability. To satisfy the requirements of both the latency and reliability, we formulate a constrained computing resource allocation optimization problem to minimize the total processing latency subject to the BS DoS probability. To efficiently solve the optimization, we design a constrained deep reinforcement learning (DRL) algorithm, which satisfies both latency and DoS probability requirements by introducing an additional critic neural network. The proposed constrained DRL further solves the issue of high input dimension by incorporating feature engineering technology.
Simulation results validate the effectiveness of our approach in achieving reliable and low-latency NS service-provisioning in the considered blockchain-secured wireless network.
\end{abstract}


\section{Introduction}
Network slicing (NS) technology is a promising technology that can support vertical industries by providing customized services using unique physical infrastructure~\cite{Survey_NS_iot}. However, potentially malicious base stations (BSs) in the communication network may jeopardize the NS service-provisioning by data tampering~\cite{NS_secure_Magezine}. Blockchain can serve as a reliable intermediary connecting physical infrastructure and requesting users, enhancing the security of NS service provisioning~\cite{DBNS_magzine, BCNS_Netletter, ICC_NSBchain}. Existing literature has shown that well-evaluated trusted BS reputations can further accelerate the effectiveness in identifying malicious BSs in blockchain networks~\cite{Xin_DBC_IoTJ, Xin_BCDRL}. 
However, how to efficiently achieve reliable NS with a minimum processing service-provisioning latency in a blockchain-secured wireless network remains open to discuss.

Existing work has applied dynamic resource allocation to improve the resource management efficiency in wireless communications~\cite{JSAC_vertical_urllc_auto_drive}. However, the diversity in customized NS service-provisioning significantly varies the physical resource holding times in different network slices~\cite{NS_concept_NGMN}, leading to highly complex sequential decision-making problems for NS resource allocation~\cite{NS_ML}. Reinforcement learning (RL) is a powerful tool for solving such sequential decision-making problems. In~\cite{Niyato_AI_sliceDeploy}, the authors applied an RL algorithm to maximize the resource utilization efficiency of the infrastructure provider in an NS network. The authors of~\cite{JSAC_NS_RL} jointly improved the energy efficiency and BS computation offloading performance by using an RL algorithm. In~\cite{JSAC_NS_MDP_duelingQ}, a deep reinforcement learning (DRL) algorithm is designed to maximize the NS provider revenue by scheduling physical resources in the wireless network. Since conventional DRL algorithm architecture composes only one critic neural network and one actor neural network, we note that few works have considered DRL for optimizing blockchain-enabled NS in wireless networks, considering the requirements of both reliability and latency.

In this paper, we propose to use constrained DRL to optimize the computing resource allocation for a blockchain-secured NS network
, where both reliability and low-latency requirements are satisfied. Customized NS services are provided by trusted BSs with high reputations, which are evaluated by user feedback and historical reputations.
The main contributions are summarized as follows: 
\begin{itemize}
    \item We formulate a computing resource optimization problem to minimize the blockchain and NS service-provisioning latency subject to a maximum denial-of-service (DoS) probability constraint. This optimization problem is one of the first to consider the joint computing resource requirements for the blockchain consensus and NS service provisioning with constraints on the network reliability.
    \item We consider the reliability is measured by the BS (DoS) probability constraints and mathematically re-formulate the problem as a Markov decision process (MDP). We enhance the feasibility of solving the MDP by aggregating processing latency features, resulting in a lower-dimensional MDP state.
    \item Finally, we transform the optimization problem from the primal domain to a dual domain and design a constrained DRL algorithm to optimize the joint computing resources for blockchain and NS service provisioning. 
\end{itemize}
Simulation results demonstrate the feasibility and effectiveness of our constrained DRL algorithm in satisfying both NS reliability and low-latency requirements in a blockchain-secured wireless network.

\section{System Model}
\begin{figure}[t]
\centering
\centerline{\includegraphics[scale=0.42]{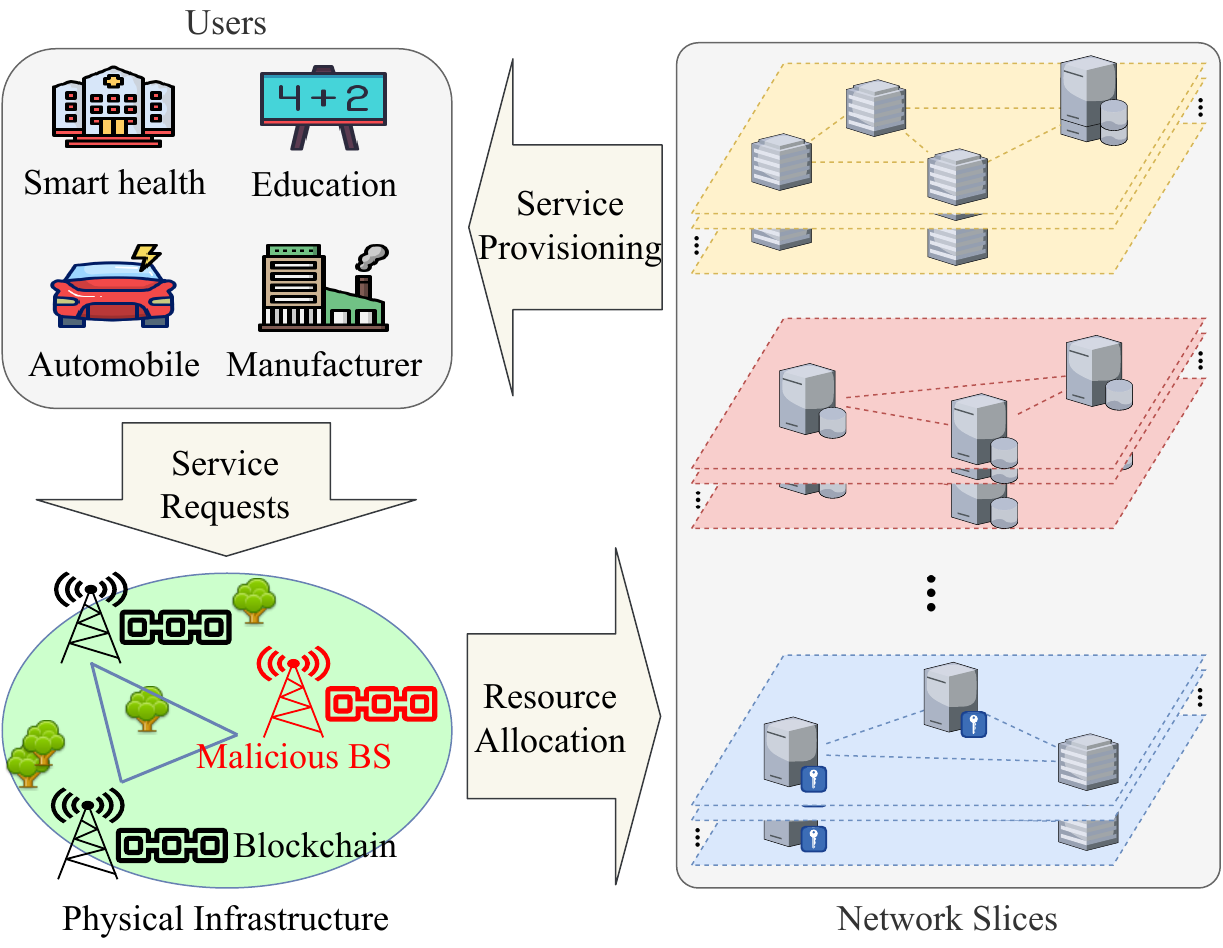}}
\caption{A blockchain-secured reliable network slicing framework for low-latency wireless communications in the presence of malicious BSs.}
\label{fig_SystemModel}
\end{figure}

Fig.~\ref{fig_SystemModel} shows our considered secure and reliable NS model, where time is discretized by time slot. We deploy $N_\text{B}$ BSs with overlapping coverage to satisfy the computing services requests from different types of users. Potentially existing malicious BSs can disrupt the computing service-provisioning process in the NS network. User requests of the same type that arrive in the same time slot are served in the same network slice. We assume that one type of request arrives in each time slot. One trusted BS is selected to process the arrived computing requests by deploying the corresponding network slice to mitigate the negative impacts caused by malicious BSs. Depending on the available resources, the selected BS applies the DRL algorithm to allocate network slices. 

The reputation of the $i$-th BS in the $t$-th time slot is updated according to
\begin{align} 
\xi_{i}(t)= 
\begin{cases} 
\xi_{i}(t-1),           & \text{if } \mathcal{D}_{i,\mathcal{K}}(t) =\varnothing \\ 
\vartheta_\text{I} I_{i}(t)  
+(1-{\vartheta_\text{I}}) {\xi}_{i}^\text{h}(t),    & \text{if } \mathcal{D}_{i,\mathcal{K}}(t) \neq\varnothing \\ 
\end{cases} 
\label{eq_Rep_update} 
\end{align}
where $I_i(t)$ is the aggregated user feedback, indicating the probability that the user requests are served by the $i$-th BS in the $t$-th time, and $\vartheta_\mathrm{I}$ is defined as the weight coefficient of reputation inference, and ${\xi}_{i}^\text{h}(t)={1}/{{\tau}_{\xi}}\sum\nolimits_{t'=t-1}^{t-{\tau}_{\xi}} \beta_{\xi}(t-t')\xi_i\left(t'\right)$ is the expected influence of historical reputations in the past ${\tau}_{\xi}$ time slots~\cite{Xin_WFIoT}. We note that eq.~\eqref{eq_Rep_update} indicates the $i$-th BS in the $t$-th time slot keeps the same value as that in the $(t-1)$-th time slot if there is no feedback received in the $t$-th time slot, otherwise evolves to an updated value.

\section{Dynamic Computing Resource Allocation}
In this section, we need to formulate the constrained optimization problem for secure and reliable NS. Next, we design a constrained DRL algorithm to optimize the resource allocation of the network slices.

\subsection{Reliable and Low-Latency Optimization}
We aim to minimize the processing latency whilst guaranteeing a given BS DoS probability constraint by optimizing the computing resources of the $i$-th BS in the $t$-th times lot, $a_i(t)$. The optimization problem is formulated as
\begin{align} 
\begin{split} 
&\min_{a_{i}(t)}\quad \mathbb{E}[\tau_{i}(t)]\\
&\begin{array}{r@{\quad}l@{}l@{\quad}l}
\text{s.t.}\quad & \mathbb{E}[{c}_{i}(t)] {\leq \epsilon_{\max}},
\label{eq: formulated_optimization} 
\end{array} 
\end{split} 
\end{align}
where $\tau_{i}(t)$ is detailed in~\eqref{eq_processing_latency} indicating the processing latency for the $i$-th BS, and ${c}_{i}(t)$ expressed in~\eqref{eq_DoS_probability} is the BS DoS probability for the $i$-th BS indicating that the assigned miner BS does not allocate any resources in the $t$-th time slot. The processing latency, $\tau_i(t)$, depends on the number of requested CPU cycles in each time slot and the allocated computing resource. Given the fixed maximum computation capacity of a BS, there is a trade-off between $\tau_{i}(t)$ and ${c}_{i}(t)$. 

\subsubsection{Processing Latency}
The processing latency corresponds to the summation latency of the processing the blockchain and the service-provisioning, i.e.,
\begin{equation}
\begin{split}
\tau_{i}(t) 
= \tau_{\text{bc},i}(t) + \tau_{\text{sp},i}(t) 
= \frac{f_{\text{bc},i}(t) + f_{\text{sp},i}(t)}{a_i(t)}, 
\quad \text{(slots)} 
\end{split}
\label{eq_processing_latency} 
\end{equation}
where $\tau_{\text{bc}, i}(t)$ and $\tau_{\text{sp}, i}(t)$ are the blockchain processing latency and the service-provisioning latency, respectively. And $f_{\text{bc},i}(t)$ and $f_{\text{sp},i}(t)$ are the corresponding requirements of the CPU cycles. Lastly, $a_{i}(t)$ (CPU cycles/slot) is the computing resources allocated in the $t$-th time slot by the {miner BS, referred to the} $i$-th BS.

We consider the service arrival process of each user request follows an independent and identically distributed (i.i.d) Bernoulli process. As such, the total number of user requests in each time slot follows a Poisson distribution, and the total number of CPU cycles required to satisfy all user requests in the $t$-th time slot can be evaluated as
\begin{equation}
    f_{\text{sp},i}(t) ={\mathds{1}\{i \in \mathcal{N}_\text{M}(t)\} \cdot {\kappa}_\text{sp}\cdot\lambda(t)\sum_{u=1}^{  {\lambda}(t)} {\ell_{u}(t)}} \quad \text{(slots)},
\end{equation}
where $\mathcal{N}_\text{M}(t)$ is the set of miner BS, ${\kappa}_\text{sp}$ (CPU cycles/byte) is the coefficient of CPU cycles required for service {provisioning}, ${\lambda}(t)$ is the Poisson distribution parameter representing the average number of user requests, and $\ell_{u}(t)$ (bytes/request) is the package size of the $u$-th requesting user.

\subsubsection{BS DoS Probability}
The instantaneous BS DoS indicator of the $i$-th BS in the $t$-th time slot is defined as
\begin{equation}
c_i(t)= \mathds{1}\{a_i(t)=0\},
\label{eq_DoS_probability}
\end{equation}
where $\mathds{1}\{\cdot\}$ is the indicator function, which equals one when no resources are allocated in the $t$-th time slot by the $i$-th BS (i.e.,  $a_{i}(t)=0$), and equals zero otherwise.

\subsection{Design of Constrained MDP}~\label{section_MDP_design}
Before apply DRL for problem solving, we first reformulate problem~\eqref{eq: formulated_optimization} as a constrained MDP. We first define the action, state, instantaneous reward and cost, and long-term reward and cost.
\subsubsection{Action}
The action to be taken in the $t$-th time slot is the computing resource of the $i$-th BS, $a_{i}(t)$ (CPU cycles/slot), shown in problem~\eqref{eq: formulated_optimization}, and is selected from the action space, 
\begin{equation}
    a_i(t) \in \mathcal{A}.
\end{equation}
We denote $\Delta f$ and $F$ as the minimum and maximum computing resources that can be allocated. If the requests are denied, $a_i(t) = 0$. Otherwise, $a_i(t)$ could be any value between ${\Delta}f$ and $F$. Thus, the action space can be described as $\mathcal{A}= \{0\} \cup \mathcal{F}$ (CPU cycles/slot), where $\mathcal{F}=[\Delta f, F]$.

\subsubsection{State}
The required CPU cycles of the requests within {each} time slot {are bounded by $f_{\text{r},\max} = \max\{f_\text{r}(t)\}$.}
Given the minimum computing resource, $\Delta f$, the maximum processing latency can be expressed as $T_{\max}={f_{\text{r},\max}}/{\Delta f}$.

We design the state of the $i$-th BS includes the states of all the computing resources allocated in the past $T_{\max}$ time slots and can be described as 
\begin{equation} 
\begin{split} 
{\boldsymbol{s}}_{i}(t)
&= \{\hat{\boldsymbol{s}}_{i}(t,t'), t' \in [t-T_{\max},t]\} \\
&= \{\hat{\boldsymbol{s}}_{i}(t,t-T_{\max}),\cdots,\hat{\boldsymbol{s}}_{i}(t-1,t),\hat{\boldsymbol{s}}_{i}(t,t)\}, 
\label{eq: state_origin} 
\end{split} 
\end{equation}
where $\hat{\boldsymbol{s}}_{i}(t,t')=[\hat{\tau}_{i}(t,t'),\hat{a}_{i}(t,t')]$ is the state of the computing resources allocated in the $t'$-th time slot, which is composed of the remaining processing latency and the computing resources allocated in the $t'$-th time slot (See Appendix~\ref{appendix_singleState}). We denote $t$ as the current time slot, and $t'$ as the time slot that the computing resource allocated, respectively. {We note} that both $t$ and $t'$ are integers, and $0 \leq t' \leq t$. {Therefore, the state space can be described as $\mathcal{S}=\{\boldsymbol{s}_i(t),  t \in [0,T_{\max}]\}$.}

\subsubsection{Instantaneous Reward and Cost}
The instantaneous reward is defined as
\begin{align} 
r_{i}(t)= 
\begin{cases} 
0,            & \text{if } a_{i}(t)=0 \\ 
-{r}_{i}^\text{a}(t),    & \text{if } a_{i}(t)\neq 0 \\ 
\end{cases} 
\label{eq: instantaneous_reward} 
\end{align} 
where {${r}_{i}^\text{a}(t)= {\tau_{i}(t)}/{\tau_{\max}}$} is the normalized processing latency when computing resource is allocated, and $\tau_{\max}$ is the maximum value of $\tau_{i}(t)$ in~\eqref{eq_processing_latency}. We denote $\overline{{\ell}}_\text{b} = \ell_\text{h}+\ell_\text{c}\overline{\lambda}$ as the average size of the blocks, $\overline{\lambda}$ as the average number of arrived requests, and $\overline{\ell}_{u_\text{r}} $ as the average size of the requests in each time slot. 

The instantaneous cost function is the BS DoS indicator function in the $t$-th time slot which is defined in~\eqref{eq_DoS_probability}. 

\subsubsection{Long-Term Reward and Cost} 
Given a policy $\mu(\hat{s}_i(t))$, the long-term discounted reward, {representing the normalized processing latency when computing resources are allocated,} is defined as
\begin{equation} 
R_{i,\mu}(t)=\mathbb{E}_{\mu}\left [\sum\limits_{\hat{t}=t}^{\infty}  {\gamma_r^{\hat{t}-t} r_{i}(t)} \right], 
\label{eq: long_term_reward} 
\end{equation} 
where $\gamma_{r}$ is the reward discount factor. The long-term discounted cost, {representing the long-term DoS probability,} is defined as
\begin{equation} 
\begin{split}
C_{i,\mu}(t)
= \mathbb{E}_{\mu}\left [\sum\limits_{\hat{t}=t}^{\infty}  {\gamma_c^{\hat{t}-t} c_{i}(t)} \right] 
=\dfrac{\mathbb{E}_{\mu}[c_{i}(t)]}{1-\gamma_{c}},
\label{eq: long_term_DoS} 
\end{split}
\end{equation} 
where $\gamma_c$ is the cost discount factor. To guarantee the requirement on the DoS probability, the long-term cost should satisfy the following constraint
\begin{equation} 
C_{i,\mu}(t) \leq 
\mathcal{E}_{\max},
\label{eq: DoS_constraint} 
\end{equation} 
where {$\mathcal{E}_{\max}={\epsilon_{\max}}/(1-\gamma_{c})$ is the maximum long-term DoS probability}, and $\epsilon_{\max}$ denotes the required threshold of the instantaneous DoS probability.

 \subsection{Reformulated Constrained MDP Problem}~\label{section_Lagran_alg}
Based on the above constrained MDP design parameters, we reformulate problem~\eqref{eq: formulated_optimization} as a constrained MDP given by
\begin{align} 
\begin{split} 
&\max_{\mu(\cdot)}\quad  R_{i,\mu}(t)\\ 
&\begin{array}{r@{\quad}l@{}l@{\quad}l} 
\text{s.t.}\quad & C_{i,\mu}(t)\leq \mathcal{E}_{\max}.
\label{eq: formulated_cmdp} 
\end{array} 
\end{split} 
\end{align} 

To solve problem~\eqref{eq: formulated_cmdp}, we utilize a constrained DRL algorithm in which the policy, $\mu(\cdot)$, and the dual variable, $\lambda_{\mathcal{L}}$, are updated iteratively. The Lagrangian function of problem~\eqref{eq: formulated_cmdp} is given by
\begin{equation} 
\mathcal{L}_{i,t}\left(\mu(\cdot), \lambda_{\mathcal{L}}\right)=R_{i,\mu}(t)-\lambda_{\mathcal{L}}\left(C_{i,\mu}(t)-\mathcal{E}_{\max}\right), 
\label{eq: Lagrangian_function} 
\end{equation} 
where $\lambda_{\mathcal{L}}$ is the Lagrangian dual variable. Problem~\eqref{eq: formulated_cmdp} can be converted to the following unconstrained problem 
\begin{equation} 
\left(\mu^{*}(\cdot),\lambda_{\mathcal{L}}^*\right)=\arg \min\limits_{\lambda_{\mathcal{L}} \geq 0} \max\limits_{\mu(\cdot)} \mathcal{L}_{i,t}(\mu(\cdot),\lambda_{\mathcal{L}}), 
\label{eq: Lagrangian_optimization} 
\end{equation} 
where $\mu^{*}(\cdot)$ and $\lambda_\mathcal{L}^{*}$ indicate the optimal policy and the optimal Lagrangian dual variable, respectively.

\subsection{Dynamic Computing Resource Allocation}
To solve the coupled MDP problem formulated in eq.~\eqref{eq: formulated_cmdp}, we decouple this problem by utilizing the PD-DDPG algorithm to find the optimal primal-dual solution~\cite{Lagrangian_PrimalDual}. We define the reward critic Q-network as $Q_{R}(\widetilde{\boldsymbol{s}},a|\theta_R)$, the cost critic Q-network as $Q_{C}(\widetilde{\boldsymbol{s}},a|\theta_C)$, the actor network as $\mu(\widetilde{\boldsymbol{s}}|\theta_{\mu})$, respectively. The corresponding target critic networks are $Q'_{R}(\widetilde{\boldsymbol{s}},a|\theta'_R)$, $Q'_{C}(\widetilde{\boldsymbol{s}},a|\theta'_C)$, and $\mu'(\widetilde{\boldsymbol{s}}|\theta'_{\mu})$. The experience replay memory buffer as $\mathcal{B}$. The specific algorithm is detailed in Algorithm~\ref{Algorithm: primal_dual_ddpg}.

\begin{algorithm}[!t] 
\algsetup{linenosize=\normalsize} \small  
\caption{Computing Resource Allocation} 
\label{Algorithm: primal_dual_ddpg} 
Randomly initialize $Q_{R}(\widetilde{\boldsymbol{s}},a|\theta_R)$, $Q_{C}(\widetilde{\boldsymbol{s}},a|\theta_C)$, and $\mu(\widetilde{\boldsymbol{s}}|\theta_{\mu})$.
Initialize $\theta'_R \gets  \theta_R$, $\theta'_C \gets  \theta_C$, and $\theta'_{\mu}\gets \theta_{\mu}$, $\lambda_{\mathcal{L}}$, $\epsilon_{\max}$, $\mathcal{B}$, $\gamma_r$, and $\gamma_c$.\\ 
{ 
{ 
Select action based on the current policy $\theta_{\mu}$ and the exploration noise $\mathcal{N}_{a}(t)$ as: 
\begin{equation} 
a(t) =\mu(\widetilde{\boldsymbol{s}}(t)|\theta_{\mu}) + \mathcal{N}_{a}(t). \nonumber 
\end{equation} 

Execute action $a_{i}(t)$ then observe reward $r(t)$, cost $c(t)$, and new state $\widetilde{\boldsymbol{s}}(t+1)$.\\ 
Store transition $\langle \widetilde{\boldsymbol{s}}(t),a(t),r(t),c(t),\widetilde{\boldsymbol{s}}(t+1) \rangle$ into $\mathcal{B}$. 
\\ 
Randomly sample a mini-batch of $M$ transition from $\mathcal{B}$: \{$\langle \widetilde{\boldsymbol{s}}_m,a_m,r_m,c_m,\widetilde{\boldsymbol{s}}_{m+1} \rangle$, $m=1,2,\cdots,M$ \}.\\ 
Set Q-targets for reward and cost temporal difference (TD): 
\begin{equation} 
\begin{split} 
y_m^{R} = r_m + \gamma_r Q'_R\left(\widetilde{\boldsymbol{s}}_{m+1}, \mu '(\widetilde{\boldsymbol{s}}_{m+1}|\theta'_{\mu})|\theta'_R\right), \\
y_m^{C} = c_m + \gamma_c Q'_C\left(\widetilde{\boldsymbol{s}}_{m+1}, \mu '(\widetilde{\boldsymbol{s}}_{m+1}|\theta'_{\mu})|\theta'_C\right).
\nonumber 
\label{eq: TD_targets} 
\end{split} 
\end{equation} 

Update reward and cost critic Q-networks by minimizing the losses: 
\begin{equation} 
\begin{split} 
\Delta_R &= \dfrac{1}{M} \sum\limits_{m}^{}\left(y_{m}^{R} - Q_R (\widetilde{\boldsymbol{s}}_{m},a_{m}|\theta_R)\right)^2,
\\
\Delta_C &= \dfrac{1}{M} \sum\limits_{m}^{}\left(y_{m}^{C} - Q_C (\widetilde{\boldsymbol{s}}_{m},a_{m}|\theta_C)\right)^2. 
\nonumber 
\label{eq: critic_q_networks} 
\end{split} 
\end{equation} 

Update the actor policy $\theta_{\mu}$, in the primal domain, with sampled policy gradient descent: 
\begin{equation} 
\begin{split} 
\nabla_{\theta_{\mu}} \mathcal{L} 
= & \dfrac{1}{M} \sum\limits_{m}^{}   
\nabla_{\theta_{\mu}} 
(Q_R\left(\widetilde{\boldsymbol{s}}_m, \mu(\widetilde{\boldsymbol{s}}_m|\theta_{\mu})|\theta_R\right) 
\\
& - \lambda_{\mathcal{L}} Q_C(\widetilde{\boldsymbol{s}}_m,\mu(\widetilde{\boldsymbol{s}}_m|\theta_{\mu})|\theta_C)).  
\nonumber 
\label{eq: primal_actor_policy_gradient} 
\end{split} 
\end{equation} 

Calculate the gradient of dual variable $\lambda_{\mathcal{L}}$: 
\begin{equation} 
\nabla_{\lambda_{\mathcal{L}}} \mathcal{L} 
= \dfrac{1}{M} \sum\limits_{m}^{}\left(Q_C(\widetilde{\boldsymbol{s}}_m,\mu(\widetilde{\boldsymbol{s}}_m|\theta_{\mu})|\theta_{C})-{\mathcal{E}_{\max}}
\right).
\nonumber 
\label{eq: dual_gradient} 
\end{equation} 

Update the dual variable, $\lambda_\mathcal{L}$, in the dual domain, with sampled dual gradient ascent: 
\begin{equation} 
\lambda_{\mathcal{L}} \gets \max\{0, \lambda_{\mathcal{L}}+\alpha_{\lambda_\mathcal{L}}\nabla_{\lambda_{\mathcal{L}}} \mathcal{L}(\theta_{\mu},\lambda_{\mathcal{L}})\}. 
\nonumber 
\label{eq: update_dual_variable} 
\end{equation} 

Update target networks with $\varphi$: 
\begin{equation} 
\begin{split} 
\theta'_{R}       &\gets \varphi \theta_{R}      + (1-\varphi)\theta'_{R}, \quad
\theta'_{C}       \gets \varphi \theta_{C}      + (1-\varphi)\theta'_{C}, 
\\
\theta'_{\mu} &\gets \varphi \theta_{\mu} + (1-\varphi)\theta'_{\mu}. 
\label{eq: update_targets} 
\nonumber 
\end{split} 
\end{equation} 
} 
} 
\end{algorithm} 


To improve the training efficiency of the constrained DRL algorithm, we propose to reduce the dimension of the state space of the constrained MDP defined in eq.~\eqref{eq: state_origin}  by extracting key features to achieve a lower dimension state for training. Specifically, we extract remaining processing latency, $\hat{\tau}_{i}(t,t')$, of all the allocated computing resources in the $i$-th BS to be a normalized sum, i.e.,
\begin{equation} 
\begin{split} 
\rho_{i}(t) 
=\dfrac{\tau_{i,\min}(t)}{\tau_{\max}}
= \dfrac{{\dfrac{1}{F}\left(\sum\limits_{t'=t-{T_{\max}}}^{t} \hat{\tau}_{i}(t,t')\right)} } 
{\tau_{\max}},
\label{eq: workload} 
\end{split} 
\end{equation} 
where $\tau_{i,\min}(t)$ is the remaining processing latency of all the allocated computing resources if they are processed with the maximum computing resource, $F$. Based on eq.~\eqref{eq: workload}, the lower dimension state is re-designed as
\begin{equation} 
\begin{split} 
\widetilde{\boldsymbol{s}}_{i}(t)=\left\{\dfrac{f_{i,\text{a}}(t)}{F},\rho_{i}(t)\right\}, 
\label{eq: feature_normalization} 
\end{split} 
\end{equation} 
where $f_{i,\text{a}}(t)$ is the available computing resources of the $i$-th BS in the $t$-th time slot.

\begin{table}[t] 
\renewcommand\arraystretch{1} 
\caption{Key Simulation Parameters} 
\centering 
\begin{tabular}{m{5.1cm} l}
\toprule 
\toprule 
\textbf{Simulation parameters}&\textbf{Values}\\
\toprule 
Number of overall BSs in the network $N_\text{B}$                                                    &10\\ 
\hline 
Computation capacity of the BSs $F$                                                                  &1.6 G CPU cycles/slot\\
\hline
Minimum computing resource allocation $\Delta f$                                                     &0.01 G CPU cycles/slot\\
\hline
Request arrival rate $\bar{\lambda}$                                                                 &$\textrm{Poisson}$($10^3$)\\ 
\hline 
Size of request by the $u$-th user ${\ell}_{u}(t)$                                                   &Uniform$(1, 10)$ KB\\
\hline 
Coefficient of reputation $\vartheta_\mathrm{I}$                                                     &0.2\\ 
\hline 
Coefficient of CPU cycles for services ${\kappa}_\text{sp}$                                          &330\\
\hline 
Threshold of DoS probability $\epsilon_{\max}$                                                       &2 \%          \\ 
\hline
Discount factors of rewards $\gamma_\text{r}$ , $\gamma_\text{c}$                                                                     &0.95\\ 
\hline
Learning rates of critic NNs $\alpha_\text{r}$, $\alpha_\text{c}$                                              &$5\times 10^{-4}$\\ 
\hline 
Learning rate of actor NN $\alpha_\text{a}$                                                                    &$2\times 10^{-4}$\\ 
\hline 
Learning rate of dual variable $\alpha_{\lambda_\mathcal{L}}$                                                  &$0.1$\\ 
\hline 
Mini-batch size $M$                                                                                            &512\\ 
\bottomrule 
\bottomrule 
\end{tabular} 
\label{tab: Simulation Parameters} 
\end{table} 

\begin{figure}[t] 
\centering 
{\includegraphics[height=6.5cm, width=9cm]{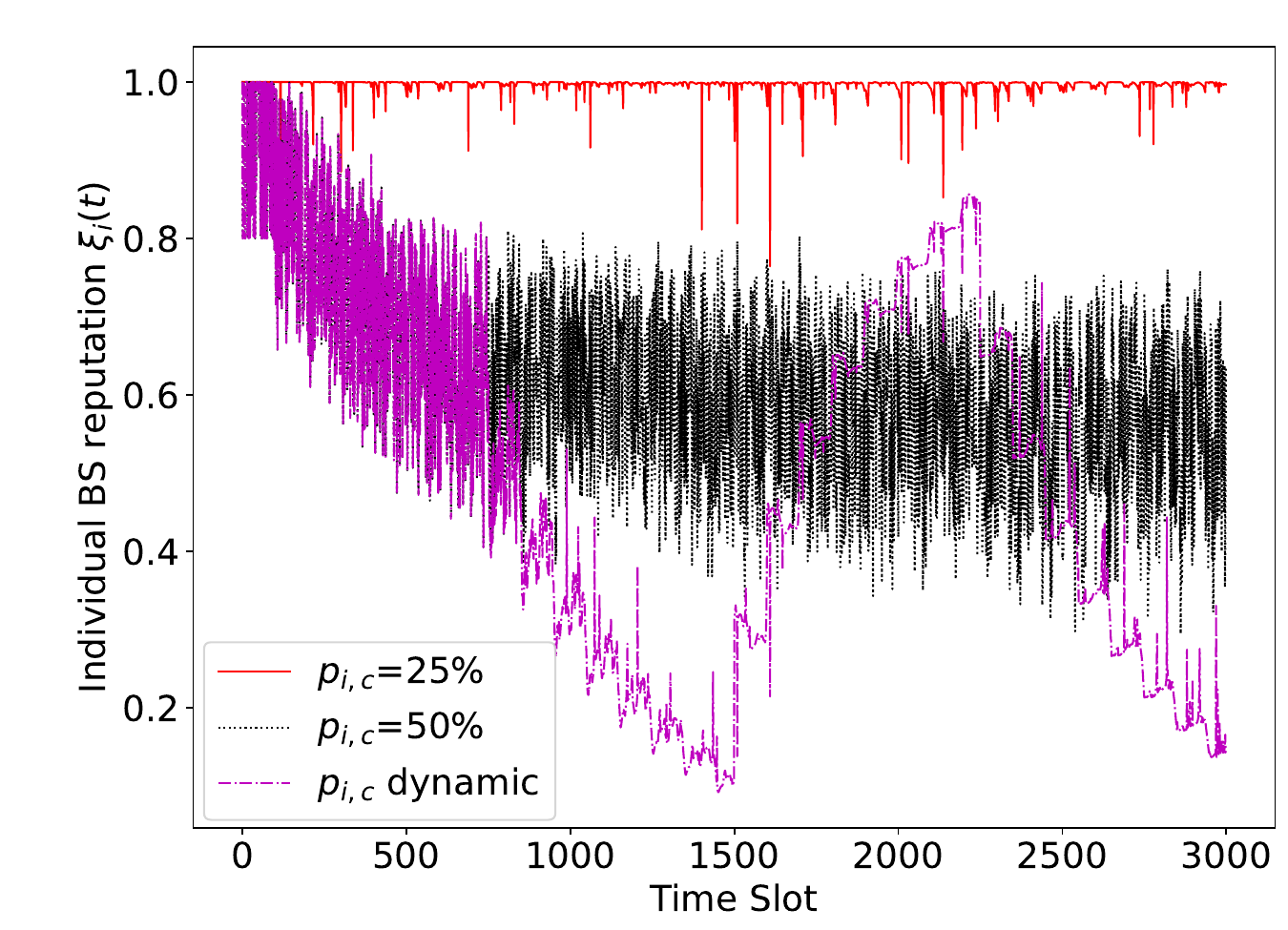}}
\caption{Real-time reputation tracking of the $i$-th BS with different probabilities of DoS feedback from the users.}
\label{fig: attack_real_time_identify} 
\end{figure}

\section{Performance Evaluation}
We present numerical simulations to evaluate the performance {and analyze the empirical convergence} of our proposed BC-DRL solution using Google TensorFlow embedded in a Python platform. We consider a total of 10 BSs that are initialized as non-malicious and have the maximum reputations equal to one. The transmission rates among all the BSs are assumed to be $W$=10Gbps. The action exploration noise in our DRL simulations follows an Ornstein-Uhlenbeck process. We note that the computing resources of the committee BSs are all considered equal to $a_i(t)$ in the $t$-th time slot, since the serving BS is randomly assigned from the committee in each time slot. Unless otherwise mentioned, the simulation parameters are summarized in Table~\ref{tab: Simulation Parameters}.

\begin{figure}[t]
\centering 
\subfigure[Normalized processing latency when resources are allocated.]
{\includegraphics[height=6.5cm, width=9cm]{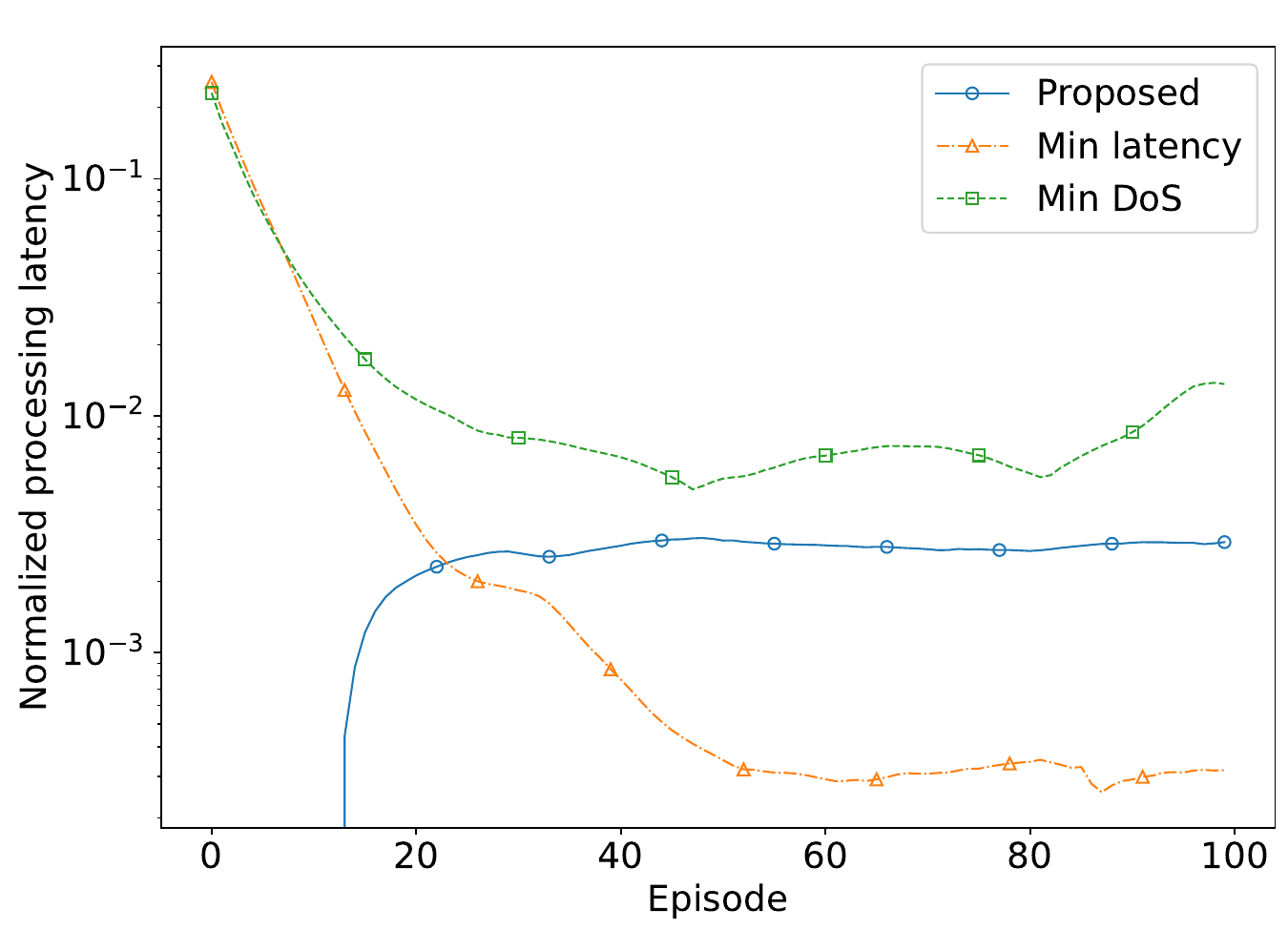}~\label{fig_train_resource_allocation_reward}} 
\subfigure[Average DoS probability in the long-term.]
{\includegraphics[height=6.5cm, width=9cm]{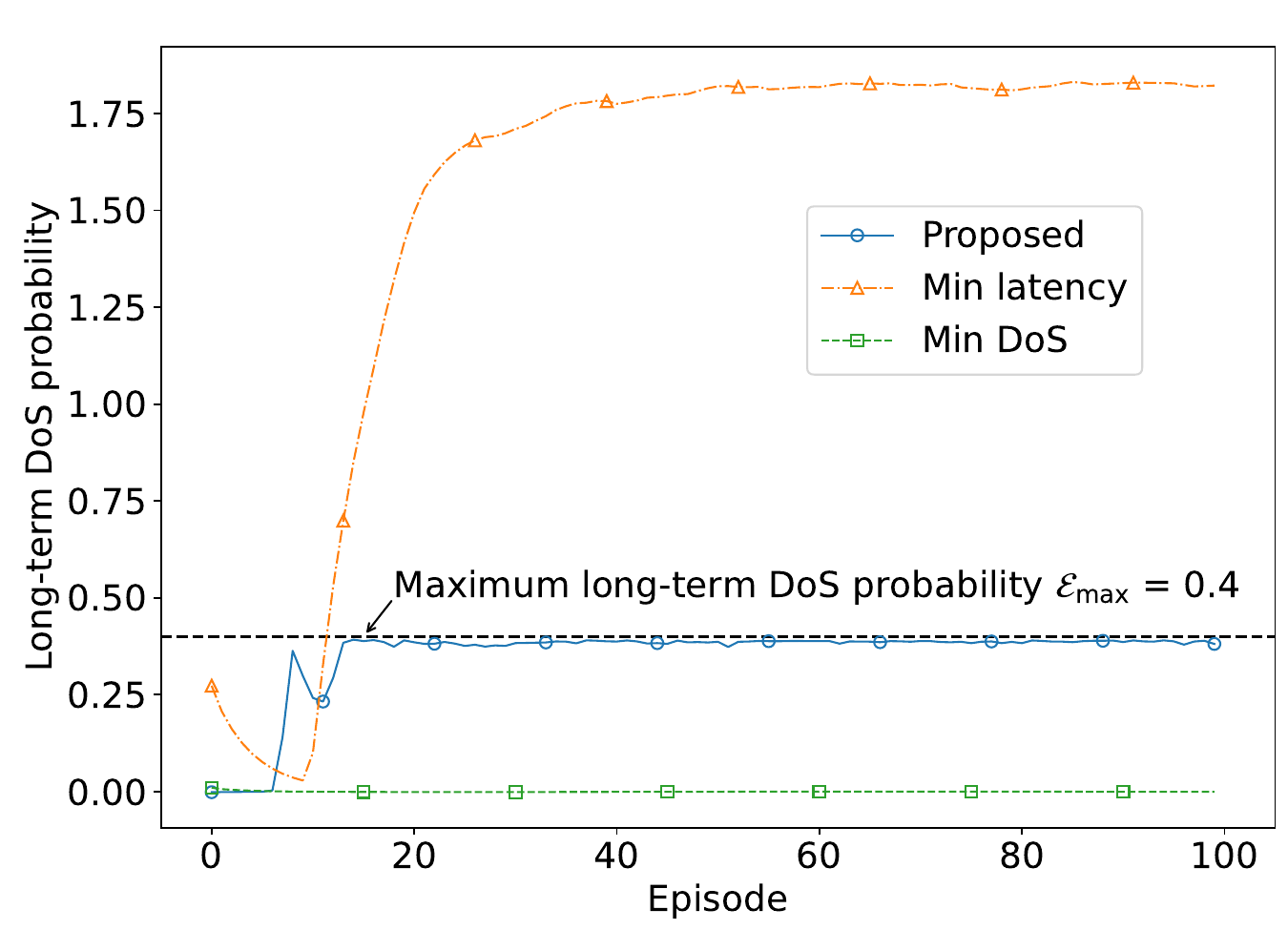}~\label{fig_train_resource_allocation_cost}} 
\caption{Processing latency and DoS probability when resources are allocated with different DRL algorithms. Our proposed algorithm achieves minimum processing latency with a satisfactory DoS probability.}
\label{fig_training_results_resource_allocation}
\end{figure}
\begin{figure}[t]
\centering 
\subfigure[Normalized processing latency when resources are allocated.]  
{\includegraphics[height=6.5cm, width=9cm]{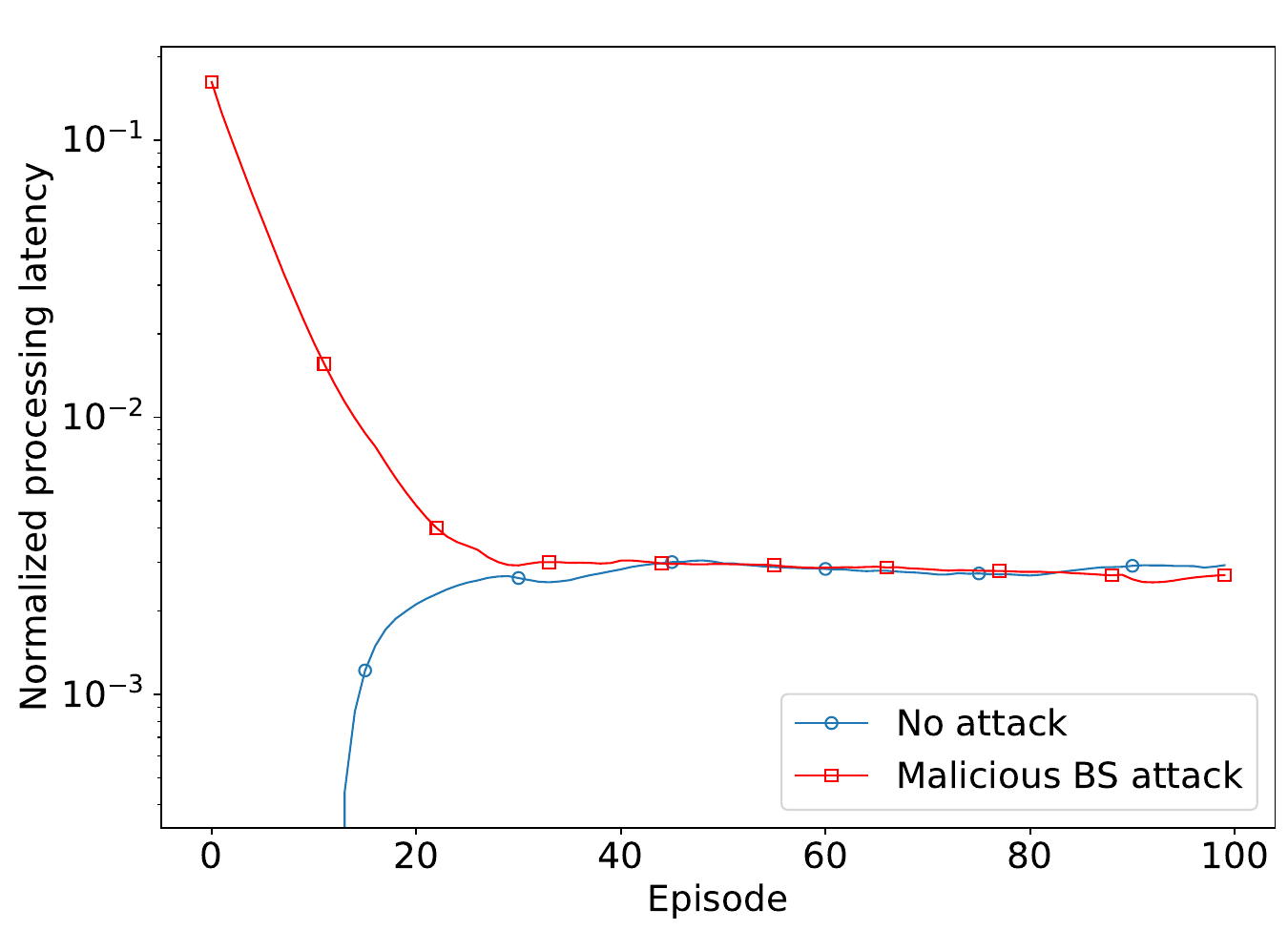}\label{fig: Training_results_reward_attack}} 
\subfigure[Average DoS probability.]
{\includegraphics[height=6.5cm, width=9cm]{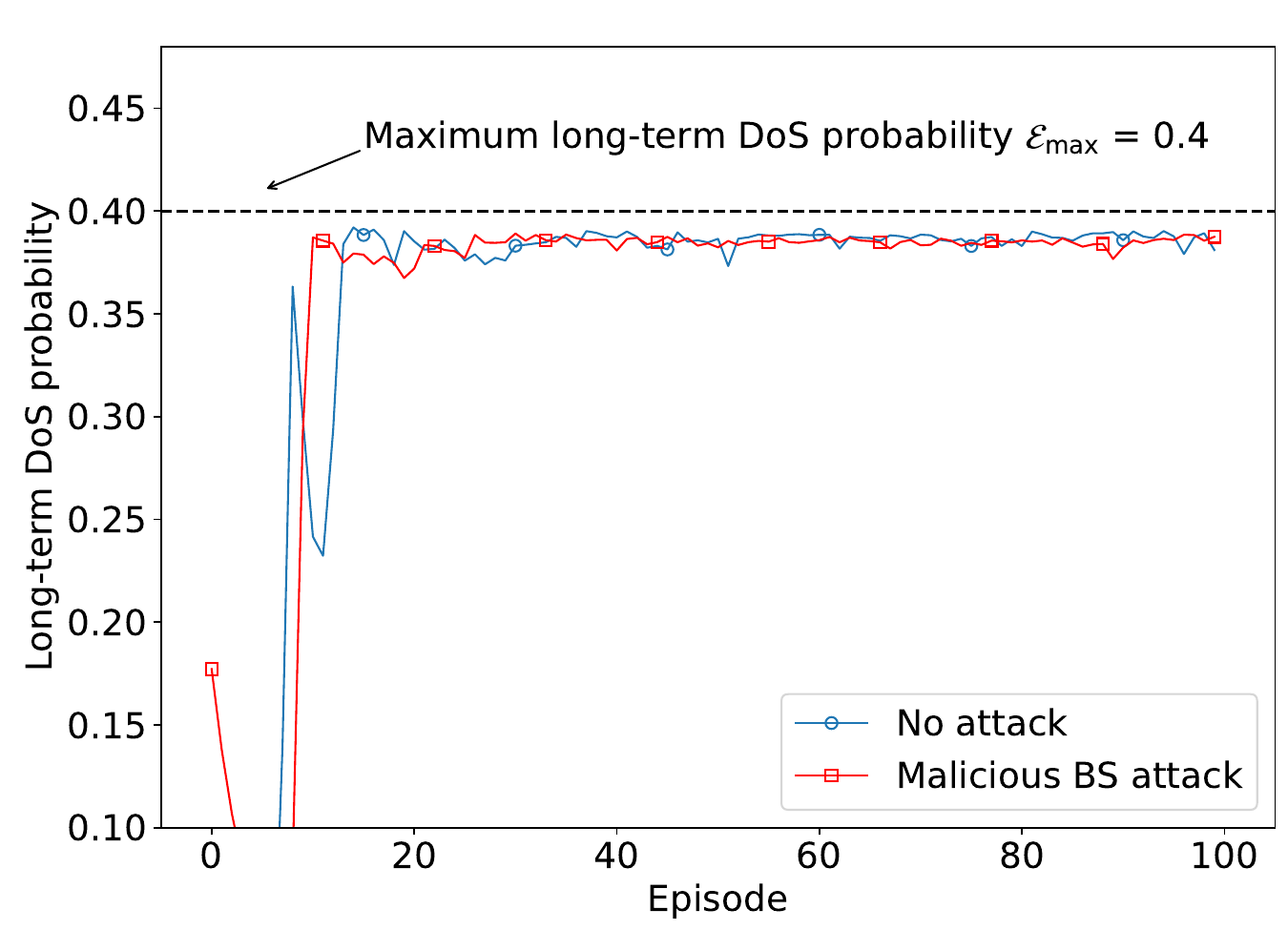}\label{fig: Training_results_cost_attack}} 
\caption{Performances of BC-DRL solution with and without {malicious BS attacks}.} 
\label{fig_training_results_attack}
\end{figure}

Fig.~\ref{fig: attack_real_time_identify} examines the tracking ability of the designed reputation evaluation method. In this figure, we consider that the feedback DoS indicators from the users for the $i$-th BS takes the value of ``1'' with probability either $p_{i,c}$= $0$, $25\%$, $50\%$, or dynamically. Fig.~\ref{fig: attack_real_time_identify} demonstrates that when $p_{i,c}$=$25\%$, the reputation of the $i$-th BS jitters at 1 and below. In contrast, when the $i$-th BS has a high level of DoS user feedback where $p_{i,c}$=$50\%$, then its reputation quickly diminishes and jitters at approximately 0.6. Lastly, when the $i$-th BS has a fluctuating level of DoS user feedback where $p_{i,c}$ changes over time, the corresponding reputation also changes accordingly. We can observe from this figure that the reputations can be tracked dynamically, ensuring the malicious BSs with high DoS user feedback will be quickly removed from the trustful committee.

In Fig.~\ref{fig_training_results_resource_allocation}, we compare our constrained DDPG DRL solution with benchmark unconstrained DDPG DRL solutions aimed at minimizing either the processing latency or DoS probability. Figs.~\ref{fig_train_resource_allocation_reward} and~\ref{fig_train_resource_allocation_cost} highlight a fundamental performance trade-off where the min latency solution leads to an intolerable DoS probability, whilst the min DoS probability solution results in a lengthy processing latency and an unnecessarily low DoS probability. In contrast, our proposed constrained DRL solution can achieve a significantly reduced processing latency while maintaining a satisfactory DoS probability, as determined by the specified maximum long-term cost.

Fig.~\ref{fig_training_results_attack} shows the training results of BC-DRL with and without malicious BS attacks. To investigate the impact of malicious BS attacks on processing latency and DoS probability, we include three malicious BSs in our simulation. Each BS randomly denies service requests from users following the Bernoulli process. We can observe that both the processing latency and the DoS probability converged to steady values after approximately 15 episodes. We observe that while BC-DRL achieves approximately the same reward and cost performance for both scenarios of with and without malicious BSs, the dual variables for the optimization converge to different values in each scenario to accurately balance the trade-off between processing latency and DoS probability.

\section{Conclusion}
We designed a reliable network slicing (NS) framework for blockchain-secured low-latency communications. Tampering attacks from the potentially existing malicious base stations (BSs) were resisted by timely evaluated BS reputations. A constrained optimization problem was formulated to minimize the processing latency subject to the constraint on the denial-of-service (DoS) probability. Low-latency was guaranteed in the objective function, whilst reliability was ensured in the DoS probability constraint. We re-formulated the problem to be a Markov decision process (MDP), and designed a constrained deep reinforcement learning (DRL) algorithm to optimize the computing resource allocation. Numerical simulation results validate the effectiveness of achieving a reliable and low-latency NS for customized wireless computing service-provisioning.

\appendices
\renewcommand{\theequation}{A.\arabic{equation}}
\setcounter{equation}{0}
\section{Derivation of Computing Resource Allocated}~\label{appendix_singleState}
Based on~\eqref{eq_processing_latency}, the processing latency of the computing resources allocated in the $t'$-th time slot is given by
\begin{equation} 
\tau_{i}(t')=\lceil \tau_{\text{bc},i}(t')+\tau_{\text{sp},i}(t') \rceil, \quad \text{(slots)} 
\label{eq: reconstuct_tau} 
\end{equation} 
where $\lceil \cdot \rceil$ is the ceiling function. In the $t$-th time slot, the remaining processing latency of the computing resources allocated in the $t'$-th time slot can be expressed as
\begin{equation}
\hat{\tau}_{i}(t,t')=(\tau_{i}(t') - (t-t'))^+, \quad \text{(slots)}.
\label{eq: remaining_time}
\end{equation}
In the $t$-th time slot, the computing resources been hold equal to the computing resources allocated in the $t'$-th time slot, and can be described as
\begin{equation}
\hat{a}_i(t,t') = a_{i}(t') \cdot\mathds{1}\{\hat{\tau}_{i}(t,t') > 0\},  \quad \text{(CPU cycles/slot)}
\end{equation}
where $\mathds{1}\{\cdot\}$ is the indicator function, which equals one when $\hat{\tau}_{i}(t,t') > 0$, and equals zero otherwise. The state of the computing resources allocated in the $t'$-th time slot is defined as its remaining processing latency and the computing resource hold by it, i.e.,
\begin{equation} 
\hat{\boldsymbol{s}}_{i}\left(t,t'\right) = \left[\hat{\tau}_{i}(t,t'),\hat{a}_{i}(t,t')\right]. 
\label{eq: queue_element} 
\end{equation} 
If the BS did not allocate any computing resources in the $t'$-th time slot, both $\hat{\tau}_{i}(t,t')$ and ${a}_{i}(t,t')$ are zeros. If the initialized processing latency $\tau_{i}(t')$ is smaller than $T_{\max}$, then $\hat{\tau}_i(t,t')$ and $\hat{a}_i(t,t')$ are also updated to zeros before $t'+T_{\max}$.

\end{document}